\documentclass[runningheads]{llncs}
\usepackage{comment}
\usepackage{cite}
\usepackage{soul,color}
\usepackage[utf8]{inputenc}
\usepackage[english]{babel}
\usepackage{cancel}
\usepackage{amsmath}
\usepackage{url}
\usepackage{array,multirow}
\usepackage{algorithm}
\usepackage{algorithmic}
\usepackage{graphicx}
\usepackage{verbatim}
\graphicspath{ {./figure/} }
\usepackage{xfrac}
\usepackage{dblfloatfix}
\usepackage{etoolbox}
\usepackage{xcolor}
\usepackage{tabularx}
\usepackage{amssymb}
\usepackage{array, boldline, makecell, booktabs}
\usepackage{enumitem}
\usepackage{diagbox}
\usepackage{afterpage}

\usepackage{caption}

\usepackage{microtype}
\usepackage{booktabs}
\usepackage[format=hang,font={small,bf}]{caption}
\usepackage{color,soul}
\setul{0.5ex}{0.3ex}
\definecolor{Green}{rgb}{0,1,0}
\setulcolor{Green}
\usepackage[bottom]{footmisc}

\usepackage{nccmath}
\usepackage{mathtools}

\begin{document}
\title{Co-Evolutionary Defence of Active Directory Attack Graphs via GNN-Approximated Dynamic Programming}
\titlerunning{Co-Evolutionary Defence of Active Directory Attack Graphs}

\author{
Diksha Goel\inst{1,2} \and
Hussain Ahmad\inst{3} \and
Kristen Moore\inst{1,2} \and
Mingyu Guo\inst{3} 
 }
\authorrunning{Goel et al.}
%
\institute{CSIRO’s Data61, Clayton, Australia \and
Cyber Security Cooperative Research Centre (CSCRC), Joondalup, Australia \\
\email{\{diksha.goel, kristen.moore\}@data61.csiro.au} \and
The University of Adelaide, Adelaide, Australia \\
\email{\{hussain.ahmad, mingyu.guo\}@adelaide.edu.au}}
%






\maketitle 

\vspace{-0.1in}
\begin{abstract}

Modern enterprise networks are increasingly built upon Active Directory (AD) systems for identity and access management, but this centralisation exposes the organisation’s control infrastructure, offering adversaries a single point of entry to compromise high-value assets. While existing AD defence approaches often assume non-adaptive attacker behaviour, real-world adversaries dynamically adapt their strategies to bypass defences, rendering such approaches brittle and ineffective. To address this limitation, we model the attacker–defender interaction in AD environments as a Stackelberg game between an adaptive attacker and a proactive defender. We propose a co-evolutionary defence framework that integrates Graph Neural Network–Approximated Dynamic Programming (GNNDP) to model attacker decision-making across large, structurally complex AD graphs, and Evolutionary Diversity Optimization (EDO) to synthesise diverse and resilient blocking strategies. To ensure scalability, we introduce a Fixed-Parameter Tractable (FPT) graph reduction technique that preserves strategic structure while reducing computational complexity. Our co-evolutionary loop jointly refines attacker and defender policies, improving generalisation to realistic adversarial patterns and avoiding premature convergence. Experiments on synthetic AD graphs demonstrate that our framework achieves near-optimal performance (within 0.1\% of optimality on r500) and consistently outperforms baselines on larger graphs (r1000 and r2000). These results validate the effectiveness of co-evolutionary training, combined with structure-aware learning, as a scalable and adaptive approach to securing enterprise environments against sophisticated threats.


\end{abstract}

\section{Introduction}
Enterprise networks are increasingly targeted by sophisticated adversaries who exploit configuration flaws and access dependencies to escalate privileges and compromise high-value assets~\cite{cao2020survey, nandi2016interdicting}. Defenders, meanwhile, must respond under uncertainty and with limited resources, making proactive, strategic countermeasures essential. These challenges are amplified in complex enterprise infrastructures, where hierarchical access and interdependent components expand the attack surface~\cite{judijanto2023edge, malatji2023management, goel2024machine, chopra2024chatnvd}.

A particularly critical target in this landscape is \textit{Microsoft Active Directory (AD)}, the central identity and access management system for Windows-based networks~\cite{Goel2022defending, goel2024optimizing, goel2023enhancing}. Over 90\% of Fortune 1000 companies rely on AD to govern authentication, authorisation, and access control~\cite{microsoft2023}, making it a high-value target. According to Enterprise Management Associates (EMA 2021), more than half of surveyed organisations experienced AD-related breaches. These breaches often follow an “identity snowball” pattern, where attackers pivot from low-privilege accounts toward the highly privileged \textit{Domain Admin (DA)} node via lateral movement~\cite{dunagan2009heat}. AD environments can be represented as attack graphs, with nodes denoting entities (e.g., users, machines, groups) and edges representing access relationships. Tools such as \textsc{BloodHound}~\cite{bloodhound} are used to analyse AD graphs by identifying shortest paths to high-value targets, but they assume static attacker strategies (these strategies assume fixed, precomputed paths that remain unchanged despite defensive actions) and fail to model adaptive behaviour. In practice, attackers frequently adapt to blocked or failed paths by exploiting alternate paths, credentials, underused privileges, and overlooked misconfigurations. Prior work for defending AD graphs has explored \textit{edge-blocking} techniques that selectively block high-impact edges. However, these methods often assume fixed attacker strategies and lack mechanisms to model dynamic adaptation, limiting their robustness and generalisability to real-world threats.

\textit{In this paper, we address this gap by proposing a scalable, structure-aware co-evolutionary defence framework that models the interaction between an adaptive attacker and a proactive defender as a Stackelberg game over AD graphs. The attacker aims to maximise their probability of reaching the DA node, while the defender, operating under a constrained budget, strategically blocks key edges to minimise attacker success.}

Our approach jointly learns attacker and defender strategies through a \textit{\textbf{unified co-evolutionary loop}} that captures both structural complexity and adaptive behaviours. To model realistic attacker behaviour, we formulate attacker decision-making as a Markov Decision Process (MDP), where each edge is associated with success, failure, and detection probabilities. The attacker starts from one of several \textit{entry nodes} and adapts its strategy based on edge outcomes. The defender, in turn, must anticipate these adaptive behaviours and deploy robust, and effective countermeasures. Unlike prior approaches~\cite{guo2021practical, dunagan2009heat, guo2024limited, ngo2024optimizing, guo2023scalable}, our method supports dynamic attacker adaptation and generalises to structurally complex AD environments. To enable scalable and effective defence of AD graphs, we introduce three key components: (1) a \textbf{\textit{Graph Neural Network–Approximated Dynamic Programming (GNNDP)}} model that approximates the attacker’s strategy using graph-aware learning, (2) an \textbf{\textit{Evolutionary Diversity Optimization (EDO)}} strategy that synthesises diverse and high-impact blocking plans, guided by GNNDP as a fitness oracle, and (3) a \textbf{\textit{Fixed-Parameter Tractable (FPT)}} graph reduction method based on \textit{Non-Splitting Paths (NSPs)}~\cite{guo2023scalable}, which preserves critical decision points while significantly reducing computational overhead.

This co-evolutionary loop enables dynamic policy refinement between attacker and defender strategies through adversarial interplay. EDO generates an evolving set of defensive configurations, persistently challenging the attacker model. In parallel, GNNDP retrains across these varying defensive configurations, generalising attacker behaviour under shifting constraints. This mutual adaptation process prevents convergence to overfitted policies, enhances robustness against distributional shifts in attacker–defender dynamics, and scales to complex, large-scale Active Directory graphs. Our results demonstrate that this co-evolutionary training yields defensive strategies that are resilient to adaptive adversaries and effective across large-scale AD structures in enterprise environments.\\
\noindent We make the following key contributions:

\begin{itemize}
\vspace{-0.11in}
\item \textit{\textbf{Attacker policy.}} We formulate the attacker’s behaviour as an MDP and propose GNNDP, a GNN-approximated dynamic programming method, enabling scalable and structure-aware computation of attack policies over large AD graphs.

\item \textit{\textbf{Defender policy.}} We propose an Evolutionary Diversity Optimization (EDO)-driven defence approach to discover diverse and robust edge-blocking defences, guided by GNNDP as a fitness oracle.

\item \textit{\textbf{Attacker–Defender co-training.}} We develop an integrated co-training mechanism where attacker and defender strategies are iteratively refined, capturing dynamic behaviours on both sides and avoiding policy stagnation.

\item \textit{\textbf{Experimental analysis.}} We conduct extensive experiments on synthetic AD graphs (r500, r1000, r2000), showing that our method achieves near-optimal performance (within 0.1\% of the optimum on r500) and consistently outperforms baselines on larger graphs, demonstrating the effectiveness of the proposed defence.
\end{itemize}

\section{Related work}
\vspace{-0.09in}
Existing approaches to defending AD graphs have mainly focused on edge interdiction and adversarial path disruption. However, they often impose restrictive assumptions that hinder scalability, adaptability, and realistic modelling of attacker behaviour.

Guo et al.\cite{guo2021practical} studied the edge interdiction problem, proposing a GNN-based model that captures attacker decision-making at local points to maximise attacker's expected shortest path lengths. However, their approach assumes that once a path is selected, the attacker proceeds without adapting to failures or blocked edges, limiting its ability to model realistic dynamic behaviours. Expanding on this, Guo et al.\cite{guo2023scalable} proposed scalable edge-blocking algorithms by exploiting the tree-like structures and short attack paths commonly found in AD graphs. While improving computational efficiency, their method assumes a non-adaptive attacker and heavily relies on structural simplifications, limiting its applicability to real-world AD environments that feature dense cycles and evolving attacker strategies. Zhang et al.\cite{Yumeng23:Near} introduced a dual-oracle defence framework that improves evaluation against industry baselines; however, their model also assumes that attackers follow fixed paths after defence deployment, without adapting to changing defensive configurations. Similarly, Goel et al.\cite{Goel2022defending} integrated neural networks with evolutionary algorithms to optimise edge-blocking policies. While effective for small graphs, their approach relied on simple feedforward neural networks that lacked the structural awareness necessary to capture the hierarchical and relational complexity intrinsic to AD graphs, reducing their generalisability across network topologies. Ngo et al.~\cite{quang,ngo2024optimizing} explored dynamic AD defence through honeypot placement and decoy deployment strategies. Although valuable for node-level deception and dynamic decoy placement, their work does not address the critical edge-blocking problem central to disrupting privilege escalation pathways. Moreover, they assume static attacker objectives and do not model adaptive adversary capable of rerouting in response to defences.

\textit{Our approach significantly advances AD defence by overcoming limitations of prior work. Unlike existing neural network-based defences~\cite{Goel2022defending} that struggle to model relational dependencies and generalise across AD topologies, we formulate attacker behaviour as a MDP and approximate it via a graph-aware GNN model. This enables scalable, structure-sensitive reasoning over complex AD graphs. In contrast to static shortest-path assumptions and local heuristic methods~\cite{guo2021practical}, our attacker model supports dynamic rerouting and policy refinement in response to evolving defensive configurations. Moreover, by integrating EDO with GNN-driven dynamic programming, we establish a co-evolutionary training loop where attacker and defender strategies mutually adapt. This synergy produces near-optimal results on small graphs and scales robustly to large enterprise environments. Together, these advancements provide a realistic, adaptive, and scalable framework for defending Active Directory infrastructures, representing a significant advancement beyond existing static or locally optimised approaches.}

\section{Problem Formulation}
\vspace{-0.09in}
We model the AD environment as a directed graph $G = (V, E)$, where $n = |V|$ denotes the number of nodes and $m = |E|$ the number of edges. An attacker initiates the attack from one of $s$ predefined \emph{entry nodes} and aims to reach the \emph{Domain Admin (DA)} node by formulating an attack policy that maximises their probability of success. To simplify analysis, multiple DA nodes, if present, are consolidated into a single representative DA node. Each edge $e \in E$ is associated with three mutually exclusive probabilities: detection $p_d(e)$, failure $p_f(e)$, and success $p_s(e) = 1 - p_d(e) - p_f(e)$. A detection event results in immediate termination of the attack; a failure blocks traversal but allows the attacker to continue exploring; and a success grants access to the edge's destination node. The attacker selects and attempts unvisited edges, progressively expanding control over the network. The attack terminates under one of three conditions: (i) the attacker is detected, (ii) no accessible paths remain, or (iii) the attacker successfully reaches the DA node. Throughout this process, the attacker accumulates a set of \emph{secured nodes},  endpoints of successfully traversed edges, which serve as new footholds for further exploration. The defender seeks to minimise the attacker's success probability by selectively blocking up to $k$ \emph{block-worthy edges} (i.e., edges eligible for blocking), where $k$ denotes the defender’s blocking budget. Only a subset of edges are considered block-worthy based on operational and strategic significance. To capture the strategic interaction between defender and attacker, we model the problem as a \emph{Stackelberg game}. The defender commits to a blocking strategy first, anticipating the attacker’s response. The attacker, with full knowledge of the blocked edges, computes an adaptive attack policy. This hierarchical setup models proactive defence planning in enterprise networks, where defenders must anticipate adversarial adaptation rather than react passively to attacks.


\section{Proposed Co-Evolutionary Attack–Defence Framework}
\vspace{-0.09in}

\begin{figure}[t!]
 \centering \includegraphics[width=0.65\textwidth]{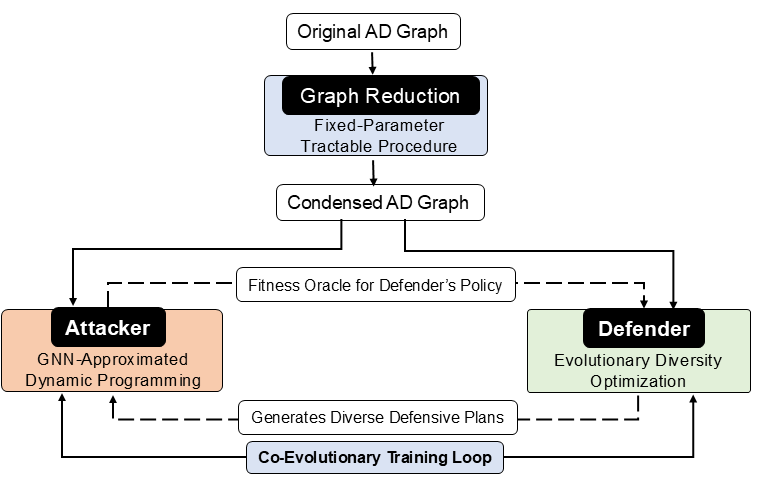}
 \caption{Co-evolutionary Framework for Defending AD Graphs.}
 \label{archi}
\end{figure}

In this section, we present our proposed attacker-defender framework, comprising three core components: (i) Fixed-Parameter Tractable (FPT) graph reduction procedure that condenses large-scale AD graphs into computationally tractable representations; (ii) Graph Neural Network-approximated Dynamic Programming based attacker policy; and (iii) Evolutionary Diversity Optimization based defensive policy for generating robust and diverse defence plans. Figure~\ref{archi} presents our proposed co-evolutionary framework for defending AD graphs through joint optimisation of attacker and defender policies.

\vspace{-0.03in}

\subsection{Graph Reduction via Fixed-Parameter Tractable Procedure}
\noindent \textit{\textbf{Condensing the AD Graph.}} We propose a Fixed-Parameter Tractable (FPT) procedure to preprocess the original AD graph into a smaller yet strategically equivalent representation. Intuitively, an AD graph can be viewed as a spanning tree augmented by $h$ \emph{feedback edges}, where $h = m - (n - 1)$ and $m$ and $n$ denote the number of edges and nodes, respectively. Within this graph, \emph{Splitting nodes} ($t$) are nodes with multiple outgoing edges, representing critical decision points for the attacker, and \emph{Entry nodes} ($s$) are nodes from which the attacker can initiate an attack. Let \textsc{SplitSet} represent the set of all splitting nodes and \textsc{EntrySet} represent the set of all entry nodes. Our FPT procedure leverages these sets to systematically condense the graph while preserving the structural properties critical to attacker–defender interactions. To facilitate this, we use the concept of {Non-Splitting Paths}~\cite{guo2021practical}:

\vspace{-0.05in}

\begin{definition} \label{def_NSP} 
\textbf{Non-Splitting Path (NSP).} A Non-Splitting Path from node $i$ to node $j$, denoted $\text{NSP}(i,j)$, is a directed path that originates at a splitting or entry node $i$, traverses intermediate nodes, each having exactly one successor, and terminates at either another splitting node or the DA node~\cite{guo2021practical}.
\end{definition}

\[
\text{NSP} = \{ \text{NSP}(i,j) \mid i \in \textsc{SplitSet} \cup \textsc{EntrySet},\ j \in \text{Successor}(i) \}
\]

We focus on those NSPs that contain at least one \emph{blockable} edge, i.e., edges that the defender is permitted to block. For each such path $\text{NSP}(i,j)$, we define its \emph{farthest blockable edge} from the source node $i$ as $bw(i,j)$. We refer to this edge as a \textbf{\textit{block-worthy edge}}, as it represents a critical point where the defender can intervene to disrupt attacker progression. The complete set of block-worthy edges is defined as:

\[
BW = \{ bw(i,j) \mid i \in \textsc{SplitSet} \cup \textsc{EntrySet},\ j \in \text{Successor}(i) \}
\]

It is important to note that a single block-worthy edge may appear in multiple NSPs. Blocking a specific path $\text{NSP}(i,j)$ consumes one unit of the defender’s budget and is realised by blocking its corresponding block-worthy edge $bw(i,j)$. This action removes the associated path from the attacker’s feasible routes to the DA node, thereby reducing their overall chance of success.

By condensing each NSP into a single {super-edge}, our FPT procedure reduces the attacker’s overall state space from $3^{|Edges|}$ to $3^{|\text{NSP}|}$. The resulting condensed graph contains ($|\textsc{SplitSet}| + |\textsc{EntrySet}| + 1$ DA) nodes and $|\text{NSP}|$ edges. This reduction preserves the critical connectivity and decision points of the original graph while significantly improving tractability for algorithmic design.

\vspace{0.1in}

\noindent \textit{\textbf{Modelling the Attacker's Problem as a Markov Decision Process.}} We formulate the attacker’s decision-making process as a Markov Decision Process (MDP), wherein each state $s$ is represented as a vector of length $|\text{NSP}|$. Each element of the state vector encodes the status of a corresponding NSP and takes
one of three values: ‘F', ‘S', ‘?'. The state vector can be represented by:

\begin{equation} \label{attacker state vector}
\text{State vector s:} \quad \underbrace{< F, ?, ?, ?, F, ?, ?, S, ?, ?, S, F >}_\text{Length = $|\text{NSP}|$}
\end{equation}

\vspace{-0.01in}
In this representation, each NSP is assigned one of the three status: ‘S’ denotes a successful traversal, indicating that the attacker has gained control of the NSP’s terminal node; ‘F’ signifies a failed attempt (e.g., due to a strong password or misconfiguration) without detection; and ‘?’ represents an unattempted path. At each decision point, the attacker selects an admissible NSP with the status ‘?’ and attempts traversal. A successful attempt updates the corresponding NSP's status to ‘S’, thereby expanding the set of nodes under the attacker's control. A failed attempt results in status update to ‘F’, but the attacker retains the ability to continue exploration from previously secured nodes. If the attacker is detected during traversal, the attack is immediately terminated. NSP marked as ‘S’ represents attacker-controlled checkpoints, from which further exploratory actions can be initiated. This iterative process continues until either the DA node is reached, the attacker is detected, or no further admissible actions remain.

\vspace{0.05in}

\noindent \textit{\textbf{State Transition Function and Action Dynamics.}}
An action\footnote{Actions available in state $s$ correspond to all unattempted NSPs that originate from nodes currently under attacker's control, i.e., the endpoints of NSPs marked as ‘S’.} $a$ executed in state $s$ leads to a set of possible successor states, denoted by $\delta(s, a)$. Upon transitioning to a new state $s'$, we first check whether $s'$ has already been encountered\footnote{Previously encountered states are stored to eliminate redundant computation; for each such state, the set of admissible actions is precomputed.}. If the state is new, its \textit{admissible action set} $A(s)$ is determined, comprising unattempted NSPs whose source nodes are reachable from the currently secured (i.e., attacker-controlled) nodes.
Actions are admissible only if their source nodes are accessible, i.e., the attacker must have previously reached the source node via a successful NSP traversal.
\textit{While the theoretical size of the state space is exponential in the number of NSPs, reaching up to $3^{|\text{NSP}|}$ (e.g., $3^{100}$ states for a graph with 100 NSPs), this renders exact computation infeasible and motivates the need for approximation.} In practice, many of these combinations are either structurally infeasible or irrelevant due to logical constraints on attacker progression. To mitigate unnecessary computational overhead, our framework restricts exploration to reachable and strategically meaningful states that reflect plausible attacker trajectories. Starting from an initial state vector (as described in Eq.~1), the state transition function iteratively determines successor states for each admissible action, thereby tracing the set of future states the attacker may encounter while progressing toward the DA node. The attacker’s objective is to maximise their probability of reaching the DA node, and this optimisation problem can be formulated using a dynamic programming approach:
\vspace{-0.01in}

\begin{equation} \label{DP}
f(s) = \max_{a \in A(s)} \left( \sum_{s' \in \delta(s,a)} \Pr(s' \mid s, a) \cdot f(s') \right)
\end{equation}

\vspace{-0.01in}
In this formulation, $f(s)$ denotes the probability that the attacker successfully reaches the DA node from state $s$. For a given action $a \in A(s)$, the attacker may transit to a set of successor states $s' \in \delta(s, a)$, each associated with a transition probability $\Pr(s' \mid s, a)$. The attacker seeks to select the action that maximises their success probability over all possible future states. Eq.~\eqref{DP} provides a principled way for computing the optimal attacker strategy via DP; however, the exponential growth of the state space renders exact computation infeasible for large-scale AD graphs. To overcome this limitation, \textit{we propose a GNN-based approximation of the attacker's value function $f(s)$, enabling scalable and structure-aware policy learning in complex enterprise environments}.

\subsection{Attacker Modelling via GNN-Approximated Dynamic Programming (GNNDP)}
\vspace{-0.03in}
We propose an approach for approximating the attacker’s policy by training a GNN to learn the value function associated with the DP formulation. Trained GNN serves as an efficient surrogate for estimating the attacker’s success probability and acts as a fitness evaluator for the defender’s EDO-based defensive policy. By leveraging the topological structure of the AD graph, the GNN iteratively aggregates information from neighbouring nodes to generate low-dimensional node embeddings. These embeddings capture both local and global graph context, allowing the model to effectively approximate attacker’s decision-making process.

Given a blocking plan, GNN receives as input the AD graph along with corresponding edge- and node-level features, and outputs the predicted probability that the attacker will successfully reach the DA node. The blocking plan is encoded as a binary edge feature, where each edge is marked as 1 if blocked and 0 otherwise. For node features, we use the top-5 diverse shortest paths to the DA node. These paths are selected to be both distinct from one another and collectively minimal in total length. To ensure diversity, we compute the similarity $\text{Sim}$ between two shortest paths $\text{SP}_1$ and $\text{SP}_2$ as the ratio of the number of shared edges to the total number of unique edges in both paths, defined as:

\begin{equation}
\text{Sim}(SP_1, SP_2) = \frac{L(SP_1 \cap SP_2)}{L(SP_1 \cup SP_2)}
\end{equation}

\vspace{-0.15in}

\noindent where $L(\cdot)$ denotes \#edges in a path. A similarity score of 1 indicates that two paths are identical, while a score of 0 means the paths are completely disjoint. To ensure sufficient diversity among the selected paths, we retain only those sets of paths in which the similarity between any pair is below a defined threshold. We empirically selected a similarity threshold of 0.4 as it offered a reasonable balance between path diversity and model stability in preliminary experiments.

For each selected path $SP$, we compute a corresponding \textit{path value}, which quantifies the likelihood that an attacker can successfully traverse the path without being detected or failing. This value is defined as:

\begin{equation}
\text{PathValue}(SP) = \prod_{e \in SP} \left(1 - (p_d(e) + p_f(e)) \right)
\end{equation}
\vspace{-0.15in}

These path values are used as node features to enhance GNN performance in estimating the attacker's success probability.

\vspace{0.1in}

\noindent \textit{\textbf{Learning Recursive Value Functions.}} To approximate the attacker’s policy, the GNN is initially trained in a supervised manner to learn the base states of the dynamic programming (DP) formulation. Since the attacker’s objective is to reach the DA node, these base states correspond to scenarios where all NSPs (Non-Splitting Paths) leading to the DA are marked either as ‘S’ (successful) or ‘F’ (failed). If at least one NSP is marked as ‘S’, it indicates that the attacker has successfully reached the DA, and the corresponding state is assigned a value of 1. Conversely, if all such NSPs are marked as ‘F’, the attacker cannot reach the DA, and the state is assigned a value of 0. In both cases, no further state transitions are possible, as the attack concludes once the DA is reached or no viable paths remain.

For all non-terminal states, the GNN is trained to approximate the recursive relationship defined in the dynamic programming formulation (Eq.\eqref{DP}). Beginning from an initial state vector (see Eq.\eqref{attacker state vector}), we simulate attack trajectories by selecting actions from the admissible action set. At each decision point, the GNN employs a mixed strategy: with 50\% probability, it chooses the action that maximises the current estimated success (exploitation), and with 50\% probability, it selects a random action (exploration). This strategy ensures a balance between reinforcing effective behaviours and discovering new states. Each action may lead to multiple possible successor states, each associated with a transition probability. One successor state is sampled based on these probabilities, and the simulation proceeds recursively until a base state is reached. This process allows the GNN to learn the recursive structure of the value function across the state space. Let $f(s; \theta)$ denote the GNN-predicted value for state $s$, with model parameters $\theta$. Our objective is to minimise the gap between the predicted value and the recursive DP target value:
\vspace{-0.15in}

\begin{equation} \label{loss}
\mathcal{L}(\theta) = \sum_{s \in S} \left(f(s;\theta) - \max_{a \in A(s)} \sum_{s' \in \delta(s,a)} \Pr(s' \mid s, a) \cdot f(s') \right)^2
\end{equation}
\vspace{-0.15in}

Here, \( S \) denotes the set of training states, \( \mathcal{A}(s) \) is the set of admissible actions at state \( s \), and \( \delta(s, a) \) is the set of successor states resulting from taking action \( a \) in state \( s \). The function \( f(s; \theta) \) is the parameterized value function, and \( \Pr(s' \mid s, a) \) is the transition probability from state \( s \) to state \( s' \) given action \( a \). This loss function minimises the gap between the GNN's predictions and the recursive DP targets, thereby performing accurate value estimation.

To generate diverse training samples, we integrate GNN training with the EDO process. In each iteration, EDO generates new and diverse blocking plans that induce different state vectors, which are then trained and evaluated using the current GNN model. These states are then evaluated and used to update the GNN model. This continuous flow of diverse samples encourages the GNN to generalise effectively across a wide range of state configurations. By combining EDO-generated diversity with recursive training, the GNN is able to approximate the attacker’s policy with high fidelity and improved robustness.

\vspace{-0.03in}

\subsection{Defender Strategy Optimisation via Evolutionary Diversity Optimisation (EDO)}
\vspace{-0.03in}

We propose an Evolutionary Diversity Optimization (EDO)-based defence strategy that aims to reduce the attacker’s probability of reaching the DA node by generating a diverse set of blocking plans. Each blocking plan disables a subset of $k$ block-worthy edges, where $k$ denotes the defender’s budget. The trained GNN, which approximates the attacker’s behaviour, serves as a fitness function to evaluate the effectiveness of each defensive configuration. EDO continuously produces a variety of blocking plans, which not only improve defensive coverage but also serve as training data to further refine the GNN. This feedback loop enables the GNN to better approximate the attacker’s policy across a broad range of defensive scenarios. Each blocking plan is represented as a binary vector over all block-worthy edges, referred to as the \textit{defensive vector}:

\begin{equation} \label{def_state_vec}
\text{Defensive vector:} \quad \langle 0, 1, 0, \ldots, 1, 0, 1 \rangle
\end{equation}

Here, a ‘1’ indicates that the corresponding block-worthy edge is selected for blocking, while a ‘0’ indicates it is left unblocked. The length of the vector equals the total count of block-worthy edges in the AD graph, and exactly $k$ entries are set to 1 to follow the blocking budget constraint.

\vspace{1em}
\noindent \textit{\textbf{Population Evolution and Fitness Evaluation.}}
We start by generating an initial population $\text{POP}$ of randomly sampled defensive vectors, where each vector encodes a valid blocking plan that selects exactly  $k$ block-worthy edges. To evolve this population over time, we apply either a mutation or crossover operation, each with equal probability. The number of bit-level changes to apply during these operations is determined by sampling a random integer $x \sim \text{Poisson}(\lambda = 1)$, which controls the degree of variation introduced in each generation.

\textbf{\textit{Mutation:}} We randomly select an individual $p_1 \in \text{POP}$, then flip $x$ randomly chosen bits from 1 to 0, and another $x$ bits from 0 to 1, preserving the blocking budget of exactly $k$ edges.

\textbf{\textit{Crossover:}} We select two individuals $p_1$ and $p_2$ from $\text{POP}$. We identify $x$ positions where $p_1$ has a 0 and $p_2$ has a 1. Now, for these positions, we change 0s to 1s in $p_1$ and 1s to 0s in $p_2$. Similarly, we look for $x$ positions where $p_1$ has 1 and $p_2$ has 0 on those positions and change 0s to 1s and 1s to 0s.

The newly generated individual is retained in the population if its fitness, defined as the attacker’s success rate under the corresponding blocking plan, as predicted by the GNN, falls within a tolerance window around the current best score. Specifically, it must lie within the range  $[f^* - 0.1,\ f^* + 0.1]$, where $f^*$ is the best fitness value observed so far. This 0.1 margin ensures that near-optimal plans are preserved while maintaining population diversity, unless the new individual significantly improves overall diversity.

\vspace{1em}
\noindent \textit{\textbf{Edge-Wise Diversity Preservation.}} We regard population diversity as an equal representation of block-worthy edges in population. Any individual $p_i$ in population $POP$ of size $\mu$ is represented as follows: 
\begin{equation*}
 p_i = \big((bw_1;i), (bw_2;i), ..., (bw_{|BW|};i)\big), \;\;\;\;\;\;i \in \{1,...,\mu\}
\end{equation*}
For each block-worthy edge $bw_j$, $\,\,j \in \{1,...,|BW|\}$, we first determine the count of individuals where this edge is blocked. We represent the block-worthy edge count of edge $bw_j$ using $c(bw_j)$. In this way, we get block-worthy edge count vector $C(bw)$ as:
\begin{equation*}
 C(bw) = (c(bw_1), c(bw_2), ..., c(bw_{|BW|}))
\end{equation*}
We calculate the diversity vector $D$ of population without an individual $p_{i}$ as: 
\begin{equation*}
 D(C(bw)\backslash{p_i}) = C(bw) - p_i
\end{equation*}
\begin{equation*}
 D(C(bw)\backslash{p_i}) = \Big(c(bw_1)-(bw_1;i),..., c(bw_{|BW|})-(bw_{|BW|};i)\Big)
\end{equation*}

where $D(C(bw)\backslash{p_i})$ describes population diversity without individual $p_{i}$. For maximising the diversity of blocked edges, we minimise the $SortedD(C(bw)\backslash{p_i})$ in lexicographic order, where sorting is performed in descending order.
\begin{flalign*}
\text{\textit{Sorted}} D(C(bw)\backslash{p_i})= sort \Big(\big(c(bw_1)-(bw_1;i)\big), .......,\big(c(bw_{|BW|})-(bw_{|BW|};i)\big)\Big)
\end{flalign*}
To maintain diversity, we identify an individual  $q$ in the population whose removal maximises overall population diversity, formally, the one that minimises $SortedD(C(bw)\backslash{p_q})$. This individual $q$ is removed if it contributes the least to diversity and its fitness score is not near-optimal. However, if the newly created individual has the highest fitness score in the population, we instead remove the individual with the lowest fitness score. Through this selective replacement strategy, the EDO process evolves a diverse and high-quality set of defensive blocking plans, which together define the defender’s policy.

\vspace{0.05in}
\noindent \textit{\textbf{Mapping Defender Plans to Attacker State Representations.}}
To compute the attacker’s probability of reaching the DA node under a given defensive blocking plan, we transform the blocking configuration (defined in Eq.\eqref{def_state_vec}) into the corresponding attacker state vector (as described in Eq.\eqref{attacker state vector}). Specifically, for each blocked block-worthy edge $bw$, we identify the associated NSP and mark its status as ‘F’ (failed), indicating that the attacker cannot traverse that path.

\subsection{Co-Evolutionary Training of GNNDP and EDO} 
\vspace{-0.03in}

Figure~\ref{archi} illustrates our proposed co-evolutionary framework, which integrates GNNDP and EDO to iteratively refine attacker and defender strategies on AD graphs. At the outset, the GNNDP model lacks sufficient training to accurately estimate the attacker’s optimal strategy or the corresponding success probability. To bootstrap the learning process, the EDO module generates a diverse set of initial blocking strategies, each inducing a distinct attacker state. These blocking configurations are converted into attacker state vectors by marking blocked NSPs as ‘F’ (failed), while all others remain as ‘?’ (unattempted). From these initial states, the attacker’s decision-making process is simulated by identifying admissible actions, computing transition probabilities, and exploring successor states via the state transition function. The GNNDP model is then trained to approximate the dynamic programming value function using the loss defined in Eq.~\eqref{loss}. This co-evolutionary training proceeds iteratively. In each cycle, EDO produces a new batch of blocking plans, which are evaluated using the current GNNDP model. These evaluations not only inform the generation of increasingly effective and diverse defence strategies but also serve as training samples to further refine the GNNDP. As the GNNDP improves in accuracy, it provides more precise feedback, guiding EDO toward better solutions. Crucially, the diversity of blocking plans generated by EDO prevents the GNNDP from overfitting to narrow regions of the state space or converging prematurely to suboptimal strategies. Instead, it promotes generalisation across attacker behaviours that are statistically likely to arise along optimal or near-optimal trajectories. As a result, training is concentrated on high-value states, improving approximation fidelity in the most strategically critical parts of the decision space.

\vspace{-0.03in}

\section{Experimental Evaluation}
\subsection{Experimental setup}

\textit{\textbf{Computational Environment.}}
All experiments were conducted on a high-performance computing cluster with Intel Xeon Gold 6148/6248 CPUs, using one core per run. The experiments required 253.3 CPU days of training in total. To accelerate training, we parallelised workload by executing 180 trials concurrently, completing all runs within two days. All models were implemented in PyTorch.
\vspace{0.05in}

\noindent \textit{\textbf{AD Graph Datasets.}} Real-world ADs are highly sensitive and not publicly available; therefore, we used the DBCREATOR tool from BloodHound to generate three structurally realistic synthetic AD graphs - r500, r1000, and r2000, each representing networks with 500, 1000, and 2000 computers, respectively. The generated AD graphs include only the default BloodHound edge types: \texttt{AdminTo}, \texttt{HasSession}, and \texttt{MemberOf}. The resulting graph statistics are as follows: r500 (1493 nodes, 3456 edges), r1000 (2996 nodes, 8814 edges), and r2000 (5997 nodes, 18795 edges). For each graph, we first selected 40 nodes farthest from the DA and then randomly selected 20 of them as attacker entry nodes. Each edge $e$ was assigned a blocking probability based on its topological proximity to the DA, given by:
\[
\text{Blocking Probability(e)} = \frac{\text{Min hops from edge $e$ to DA}}{\text{Max hops from any edge to DA}}
\]

Prior to model training, we preprocessed each AD graph to obtain a condensed graph. The preprocessing steps included (1) merging multiple DA nodes into a single unified node, (2) removing irrelevant nodes and edges (e.g., edges originating from DA or terminating at entry nodes), and (3) compressing each NSP between decision nodes into a single edge, thereby significantly reducing graph complexity.
\vspace{0.02in}

\noindent \textit{\textbf{Correlation Analysis of Edge Features.}}
To investigate the impact of edge feature correlation on attacker success, we explore three statistical relationships between edge failure probability $p_{f(e)}$ and detection probability $p_{d(e)}$: \textit{Independent (I)}, \textit{Positive Correlation (P)}, and \textit{Negative Correlation (N)}.

\begin{itemize}
 \item \textit{Independent (I):} $p_{f(e)}$ and $p_{d(e)}$ are independently sampled from a uniform distribution over $[0, 0.2]$.
 
 \item \textit{Correlated (P/N):} Both values are drawn from a multivariate normal distribution:
 \[
 (p_{d(e)}, p_{f(e)}) \sim \mathcal{N}(\boldsymbol{\mu}, \boldsymbol{\Sigma}), \quad \boldsymbol{\mu} = [0.1, 0.1]
 \]
 \begin{itemize}
 \item \textit{Positive Correlation (P):}
 \[
 \boldsymbol{\Sigma} = \begin{bmatrix}
 0.05^2 & 0.5 \cdot 0.05^2 \\
 0.5 \cdot 0.05^2 & 0.05^2
 \end{bmatrix}
 \]
 
 \item \textit{Negative Correlation (N):}
 \[
 \boldsymbol{\Sigma} = \begin{bmatrix}
 0.05^2 & -0.5 \cdot 0.05^2 \\
 -0.5 \cdot 0.05^2 & 0.05^2
 \end{bmatrix}
 \]
 \end{itemize}
\end{itemize}

These configurations allow us to analyse how statistical dependencies between edge-level features affect the success rate of the attacker.
\vspace{0.05in}

\noindent \textit{\textbf{GNNDP Training Configuration.}} 
The proposed GNNDP model receives three inputs: (i) five node features representing the probabilities of the top-5 diverse shortest paths to the DA node, (ii) one binary edge feature indicating whether the edge is blocked, and (iii) AD graph structure. A linear encoder expands both node and edge features to 128-dimensional embeddings. The model comprises five crystal graph convolutional layers, each followed by a ReLU activation function. The output is passed through a linear projection to a single dimension, aggregated using a mean pooling layer. A sigmoid activation function is applied to produce a final prediction in the range $[0,1]$, representing the estimated probability that the attacker reaches the DA node. Training is performed using mini-batches of size 16, with the Mean Squared Error loss function and the Adam optimiser. The learning rate is set to 0.001, and each training cycle runs for 300 epochs.
\vspace{0.05in}

\noindent \textit{\textbf{EDO Training Configuration.}}
We implemented EDO algorithm to generate defensive plans under a strict blocking budget of 5 edges. EDO generates 100 blocking plans (i.e., population size) over 10,000 iterations, applying mutation or crossover operations with a probability of 0.5 each. We performed 100 iterative co-training rounds involving GNNDP model updates and EDO defence generation.
\vspace{-0.2in}

\subsection{Baselines}
To assess the performance of our proposed framework, we compare it against various attacker–defender strategy combinations:

\begin{enumerate}[leftmargin=*,noitemsep,topsep=0pt]
\item {\textit{GNNDP-EDO (Proposed).}} GNNDP is used to solve the attacker's problem, while the EDO is employed for generating the defender’s blocking strategies. The defender explores diverse defensive plans, and the fitness of each plan is computed using GNNDP.

\item {\textit{GNNDP-EDO+DP.}} The attacker’s policy is modelled using GNNDP, while EDO generates candidate defences. However, instead of relying on GNNDP to estimate the attacker’s success probability, we use exact DP to provide precise evaluations for the best blocking plan.

\item {\textit{Optimal Solution.}} It leverages DP to model the attacker and exhaustively searches all combinations of edge blockings to determine the best defence.

\item {\textit{GNNDP-SEC.}} GNNDP models the attacker’s problem, and Score-based Evolutionary Computation (SEC) is used as a defender strategy. Only high-scoring blocking plans are retained based on their fitness as evaluated by GNNDP.

\item {\textit{GNNDP-Greedy.}} GNNDP computes the attacker’s policy while the defender applies a greedy strategy. In each step, the defender blocks the most critical edge, that is, the edge whose removal leads to the greatest reduction in the attacker's success rate, as estimated by GNNDP. This process is repeated until the edge-blocking budget $k$ is reached.
\end{enumerate}

\subsection{Results} 
\vspace{-0.05in}

For the smaller AD graph r500, we evaluate our framework, GNNDP-EDO, against GNNDP-EDO+DP and the Optimal solution. In contrast, for larger graphs like r1000 and r2000, computing the success rate for all attacker states using DP is computationally infeasible. Therefore, to evaluate the performance of GNNDP-EDO on larger graphs, we compare it against GNNDP-SEC and GNNDP-Greedy. Given an attacker's strategy, the \textit{success rate} quantifies the likelihood of the attacker reaching the DA node without being detected. \textit{Time} refers to the number of computation hours required for each experimental run. As the trained GNNDP model may not always yield precise estimates of the success rate, we validate its predictions using Monte Carlo simulations, conducting 25,000 runs for the best blocking plan identified by GNNDP. Each experimental setting was run ten times on independently generated AD graph instances with varying entry nodes and block-worthy edges. Reported results represent the average across these trials to ensure robustness and generalizability.\\

\begin{table*}[t!] 
\caption{Attacker success rate on the r500 AD graph under different approaches.}
\label{r500}
\renewcommand{\arraystretch}{1.3}
\centering 
\begin{tabular}{|c|c|c|c|c|} \hlineB{2} 


 \textbf{Approach} &\textbf{I} & \textbf{P} & \textbf{N} & \textbf{Avg}\\ \hlineB{2} 

GNNDP-EDO (Proposed) & 89.74\% & 84.84\% & 84.68\% & 86.42\%\\
 GNNDP-EDO + DP & 89.78\% & 84.86\% & 84.58\% & 86.4\% \\
 Optimal solution & 89.73\% & 84.78\% & 84.45\% & 86.32\% \\\hlineB{2} 
\end{tabular} 
\end{table*}
\vspace{-0.03in}
\noindent \textit{\textbf{Results for r500.}} Table~\ref{r500} presents the attacker's success rates on the r500 AD graph under three distinct statistical distributions. The \textit{Avg} column represents the mean success rate across these distributions. The results show that the attacker’s average success rate under the proposed defence GNNDP-EDO is 86.42\% when simulated using Monte Carlo simulations (average over three distributions). When the same defence is evaluated with the exact DP method (GNNDP-EDO+DP), the success rate is 86.4\%, indicating that the GNNDP model accurately approximates the attacker’s value function, with a deviation of only 0.02\%. For comparison, the optimal defence computed through exhaustive search results in a success rate of 86.32\%. This shows that our proposed GNNDP-EDO approach performs within 0.1\% of the optimal, demonstrating its effectiveness and accuracy.

\begin{table*}[t!] 
\caption{Comparison of attacker success rates on the r1000 Active Directory graph across different defence strategies. Lower values indicate stronger defensive effectiveness.}
\label{tab:r1000}
\renewcommand{\arraystretch}{1.3}
\centering 
\begin{tabular}{|c|c|c|c|c|c|c|c|} \hlineB{2} 

\multirow{2}{*}{\textbf{Approach}}& \multicolumn{4}{c|}{\textbf{Success Rate (\%)}} & \multicolumn{3}{c|}{\textbf{Time (hours)}} \\ \cline{2-8}
 & \textbf{I} & \textbf{P} & \textbf{N} & \textbf{Avg} & \textbf{I} & \textbf{P} & \textbf{N} \\ \hlineB{2} 

GNNDP-EDO (Proposed) & \textbf{44.98}\% & \textbf{46.72}\% & 45.78\% & \textbf{45.82}\% & 39.33 & 38.99 & 36.6 \\
GNNDP-SEC & 45.63\% & 47.95\% & \textbf{44.29}\% & 45.95\% & 39.38 & 35.94 & 35.83 \\
GNNDP-Greedy & 53.67\% & 51.40\% & 48.51\% & 51.19\% & 40.42 & 34.57 & 36.28 \\ \hlineB{2} 

\end{tabular} 
\end{table*}

\begin{table*}[t!] 
\caption{Comparison of attacker success rates on the r2000 Active Directory graph across different defence strategies. Lower values indicate stronger defensive effectiveness.}
\label{tab:r2000}
\renewcommand{\arraystretch}{1.3}
\centering 
\begin{tabular}{|c|c|c|c|c|c|c|c|} \hlineB{2} 

\multirow{2}{*}{\textbf{Approach}}& \multicolumn{4}{c|}{\textbf{Success Rate (\%)}} & \multicolumn{3}{c|}{\textbf{Time (hours)}} \\ \cline{2-8}
 & \textbf{I} & \textbf{P} & \textbf{N} & \textbf{Avg} & \textbf{I} & \textbf{P} & \textbf{N} \\ \hlineB{2} 

GNNDP-EDO (Proposed) & \textbf{37.91}\% & \textbf{36.20}\% & 39.36\% & \textbf{37.82}\% & 32.33 & 28.54 & 28.88 \\
GNNDP-SEC & 38.91\% & 38.12\% & \textbf{39.03}\% & 38.68\% & 33.37 & 29.44 & 28.3 \\
GNNDP-Greedy & 44.06\% & 43.26\% & 43.45\% & 43.59\% & 31.77 & 28.84 & 29.01 \\ \hlineB{2} 

\end{tabular} 
\end{table*}

\vspace{0.02in}
\noindent \textit{\textbf{Results for r1000 and r2000.}} Table~\ref{tab:r1000} presents the attacker's success rates on the r1000 AD graphs, where the attacker’s policy is approximated using GNNDP. On the r1000 graph, the proposed EDO-based defence achieves the lowest average attacker success rate of 45.82\%, compared to 45.95\% under the SEC-based defence and 51.19\% under the Greedy defence. Table~\ref{tab:r2000} presents results for the r2000 AD graphs, and the results show that the attacker's average success rate is 37.82\% under the EDO-based defence, increasing to 38.68\% under SEC and 43.59\% under Greedy. On average, across both the r1000 and r2000 graphs, the EDO-based defence consistently outperforms the SEC-based and Greedy defences. This performance stems from EDO's ability to explore a diverse range of blocking strategies rather than converging prematurely on suboptimal configurations, enabling it to find globally robust defences. In contrast, Greedy performs the worst because it evaluates edges in isolation and greedily blocks those with the highest immediate impact, often neglecting their long-term utility in multi-step attack paths. This shortsighted approach results in poor coverage of strategic paths, enabling attackers to exploit alternatives. Under negatively correlated edge distributions, the SEC-based defence slightly outperforms EDO. This may be because SEC more effectively exploits the trade-off between failure and detection probabilities, while EDO’s diversity-driven strategy may occasionally select suboptimal edges in such structured settings.

Our results indicate that, on average, the proposed GNNDP-EDO defence provides the most effective and generalisable defence against the attacker. Moreover, the results suggest that the EDO optimisation process, approximated by the GNNDP model, effectively identifies high-quality defensive policies. This highlights the suitability of the trained GNNDP as a fitness function for EDO.

\section{Discussion}
\vspace{-0.07in}

\noindent \textbf{Summary of Findings.} The framework’s integration of GNNDP with EDO captures dynamic attacker–defender interplay, achieving near-optimal performance (within 0.1\% of the optimum on r500) and outperforming baselines on larger graphs. Its co-evolutionary loop, where EDO generates diverse blocking strategies that challenge the GNNDP-based attacker model, while GNNDP continuously refines defensive responses, prevents overfitting and ensures robust generalisation. This `arms race' dynamic closely mirrors real-world adversarial interactions and surpasses static or fixed-path defence methods that fail to counter adaptive adversaries. 

\noindent \textbf{Practical Benefits for SOC Operations.} The framework provides actionable utility for Security Operations Center (SOC) teams by automating the prioritisation of high-impact edge-blocking decisions in AD environments. It can generate a ranked list of vulnerable access paths, enabling analysts to proactively block critical relationships (e.g., \texttt{AdminTo} edges) through integration with tools such as Microsoft Defender for Identity or Splunk. The GNNDP module produces interpretable outputs, quantifying attacker path success probabilities, that support transparent decision-making, efficient resource allocation, and timely incident response. The framework is designed for operational scalability, demonstrating the ability to process enterprise-scale AD graphs (e.g., r2000 with 18,795 edges). It can also be integrated with SIEM platforms, supporting seamless incorporation into existing enterprise security workflows and enhancing operational efficiency. Beyond immediate deployment, the framework promotes proactive defence planning and compliance auditing.

\noindent \textbf{Limitations and Future Directions.} Our evaluation relies on synthetic AD graphs generated via BloodHound’s DBCreator, which offer structural realism but do not fully capture the operational complexity of production environments, such as evolving user behaviours, dynamic access policies, and diverse edge types. Additionally, structural patterns and attack paths in real-world AD deployments may differ from those in synthetic graphs, which could impact the generalisability of learned defence policies. Future work will aim to validate the framework on production AD environments, incorporating richer edge semantics and behavioural dynamics, and explore the integration of Large Concept Models (LCMs) \cite{ahmad2025future} as high-level reasoning layers to augment graph-based learning with contextual understanding of vulnerabilities, adaptive defence narratives, and automated policy explanations.

\section{Conclusion}
\vspace{-0.07in}

We proposed a co-evolutionary framework for defending Active Directory (AD) graphs by integrating GNNDP with EDO. This integration enables the joint evolution of attacker and defender strategies: the attacker refines its policy using defensive plans generated by EDO as training data, while the defender iteratively generates robust blocking configurations through GNN-guided exploration. Experimental results show that this synergy achieves near-optimal performance on smaller AD graphs and consistently outperforms baseline methods on larger ones. These findings highlight the strength of combining dynamic programming with structure-aware learning. GNNDP effectively models adaptive attacker behaviour in high-dimensional spaces, while EDO facilitates the discovery of diverse and resilient defences, avoiding premature convergence to suboptimal strategies. This co-evolutionary process fosters an “arms race” dynamic, encouraging strategic generalisation for both attacker and defender. Beyond AD defence, our approach demonstrates a broader paradigm: neural approximations combined with evolutionary search offer a scalable and effective alternative to exact methods in complex adversarial settings.

\bibliographystyle{splncs04}
\bibliography{mybib}

\begin{thebibliography}{10}
\providecommand{\url}[1]{\texttt{#1}}
\providecommand{\urlprefix}{URL }
\providecommand{\doi}[1]{https://doi.org/#1}

\bibitem{ahmad2025future}
Ahmad, H., Goel, D.: The future of ai: Exploring the potential of large concept models. arXiv preprint arXiv:2501.05487  (2025)

\bibitem{bloodhound}
{BloodHoundAD}: Bloodhound - active directory enumeration tool. \url{https://github.com/BloodHoundAD/BloodHound}, accessed: 2025-01-13

\bibitem{cao2020survey}
Cao, L., Jiang, X., Zhao, Y., Wang, S., You, D., Xu, X.: A survey of network attacks on cyber-physical systems. IEEE Access  \textbf{8},  44219--44227 (2020)

\bibitem{chopra2024chatnvd}
Chopra, S., Ahmad, H., Goel, D., Szabo, C.: Chatnvd: Advancing cybersecurity vulnerability assessment with large language models. arXiv preprint arXiv:2412.04756  (2024)

\bibitem{dunagan2009heat}
Dunagan, J., Zheng, A.X., Simon, D.R.: Heat-ray: combating identity snowball attacks using machine learning, combinatorial optimization and attack graphs. In: Proceedings of the ACM SIGOPS 22nd symposium on Operating systems principles. pp. 305--320 (2009)

\bibitem{goel2023enhancing}
Goel, D.: Enhancing network resilience through machine learning-powered graph combinatorial optimization: Applications in cyber defense and information diffusion. arXiv preprint arXiv:2310.10667  (2023)

\bibitem{goel2024machine}
Goel, D., Ahmad, H., Jain, A.K., Goel, N.K.: Machine learning driven smishing detection framework for mobile security. arXiv preprint arXiv:2412.09641  (2024)

\bibitem{goel2024optimizing}
Goel, D., Moore, K., Guo, M., Wang, D., Kim, M., Camtepe, S.: Optimizing cyber defense in dynamic active directories through reinforcement learning. In: European Symposium on Research in Computer Security. pp. 332--352. Springer (2024)

\bibitem{Goel2022defending}
Goel, D., Ward-Graham, M.H., Neumann, A., Neumann, F., Nguyen, H., Guo, M.: Defending active directory by combining neural network based dynamic program and evolutionary diversity optimisation. In: Proceedings of the Genetic and Evolutionary Computation Conference. p. 1191–1199. GECCO '22 (2022)

\bibitem{guo2021practical}
Guo, M., Li, J., Neumann, A., Neumann, F., Nguyen, H.: Practical fixed-parameter algorithms for defending active directory style attack graphs. In: Proceedings of the AAAI Conference on Artificial Intelligence. vol.~36, pp. 9360--9367 (2022)

\bibitem{guo2024limited}
Guo, M., Li, J., Neumann, A., Neumann, F., Nguyen, H.: Limited query graph connectivity test. In: Proceedings of the AAAI Conference on Artificial Intelligence. vol.~38, pp. 20718--20725 (2024)

\bibitem{guo2023scalable}
Guo, M., Ward, M., Neumann, A., Neumann, F., Nguyen, H.: Scalable edge blocking algorithms for defending active directory style attack graphs. In: Proceedings of the AAAI Conference on Artificial Intelligence. vol.~37, pp. 5649--5656 (2023)

\bibitem{judijanto2023edge}
Judijanto, L., Hindarto, D., Wahjono, S.I., et~al.: Edge of enterprise architecture in addressing cyber security threats and business risks. International Journal Software Engineering and Computer Science (IJSECS)  \textbf{3}(3),  386--396 (2023)

\bibitem{malatji2023management}
Malatji, M.: Management of enterprise cyber security: A review of iso/iec 27001: 2022. In: 2023 International conference on cyber management and engineering (CyMaEn). pp. 117--122. IEEE (2023)

\bibitem{microsoft2023}
{Microsoft Corporation}: Microsoft digital defense report 2023. Tech. rep., Microsoft (2023), \url{https://www.microsoft.com/en-us/security/security-insider/threat-landscape/microsoft-digital-defense-report-2023}, accessed: 2025-04-23

\bibitem{nandi2016interdicting}
Nandi, A.K., Medal, H.R., Vadlamani, S.: Interdicting attack graphs to protect organizations from cyber attacks: A bi-level defender--attacker model. Computers \& Operations Research  \textbf{75},  118--131 (2016)

\bibitem{ngo2024optimizing}
Ngo, H.Q., Guo, M., Nguyen, H.: Optimizing cyber response time on temporal active directory networks using decoys. arXiv preprint arXiv:2403.18162  (2024)

\bibitem{quang}
Ngo, Q.H., Guo, M., Nguyen, H.: Near optimal strategies for honeypots placement in dynamic and large active directory networks. In: The 22nd International Conference on Autonomous Agents and Multiagent Systems (2023), extended Abstract

\bibitem{Yumeng23:Near}
Zhang, Y., Ward, M., Guo, M., Nguyen, H.: A scalable double oracle algorithm for hardening large active directory systems. In: The 18th ACM ASIA Conference on Computer and Communications Security (ACM ASIACCS), Melbourne, Australia, 2023 (2023)

\end{thebibliography}

\end{document}